# Rocksalt rare-earth monoxides as electronic and magnetic materials

Tomoteru Fukumura[1,3,4,*], Satoshi Sasaki[1], Masamichi Negishi[1], Daichi Oka[2]

[1] Department of Chemistry, Graduate School of Science, Tohoku University, Sendai 980-8578, Japan

[2] Department of Chemistry, Graduate School of Science, Tokyo Metropolitan University, Tokyo 192-0397, Japan

[3] WPI Advanced Institute for Materials Research, Tohoku University, Sendai 980-8577, Japan

[4] Center for Science and Innovation in Spintronics, Tohoku University, Sendai 980-8577, Japan

*Correspondence: tomoteru.fukumura.e4@tohoku.ac.jp (T. Fukumura).

**Abstract**

Stable binary rare earth ($RE$) oxides are usually trivalent $RE$ ion sesquioxides ($RE_2O_3$), that are highly insulating and either nonmagnetic or antiferromagnetic. On the other hand, rocksalt-type divalent $RE$ ion monoxides ($RE$Os) have been scarcely synthesized owing to their metastable nature. Accordingly, their fundamental properties have not been unveiled. Recently, thin film epitaxy was successfully applied to synthesize various $RE$Os. In stark contrast with $RE_2O_3$, $RE$Os are highly electrically conducting, and exhibit superconductivity, room temperature ferromagnetism, and so on. Therefore, $RE$Os are new and simple f-electron system, promising as electronic, magnetic, and spintronic materials. In this review, their fundamental properties are introduced, and their significance and future prospects are discussed toward a new paradigm in f-electron systems.

**Rare earth oxides**

Rare earth (*RE*) elements are Sc, Y, and the lanthanoids (La−Lu). Their abundances in the earth's crust are not rare, contrary to their name, but are larger than those of noble metal elements (Ru−Ag, Os−Au). Consequently, the *RE* elements are used in a broad range of materials like permanent magnets, magnetic and optical recording media, phosphors, lasers, optical fibers, and high temperature superconductors [1]. *RE* elements are easily oxidized in air by losing all s and d valence electrons from the *RE* atom to form binary *RE* oxides with closed shell trivalent *RE* ions, usually sesquioxides $RE_2O_3$ [2,3]. The $RE_2O_3$ series are wide gap insulators, and either diamagnetic, paramagnetic, or antiferromagnetic with low Néel temperatures ($T_N$) (≤ 7 K) **(see Glossary)**. These properties originate from the localized 4f electrons without the valence electrons (if *RE* = lanthanoid). The high insulating nature and high chemical stability of $RE_2O_3$ series mean that they are applied as various kinds of sensors, insulating layers in capacitors, memories, and field effect transistors **(see Glossary)**, etc [4].

Generally, trivalent *RE* ions are very stable in *RE* compounds, while unusual divalent *RE* ions are also present in solid-phase rocksalt-type *RE* monoxides (*RE*O) in addition to *RE*O gas molecules [5] and *RE* metal complexes [6]. Divalent *RE* metal complexes have been already synthesized to explore the unusual $4f^n5d^1$ electronic configuration of *RE* ions as possible new catalysts and molecular magnets [6,7]. On the other hand, the solid-phase *RE*Os have been rarely synthesized despite the very simple rocksalt structure, unlike the divalent *RE* metal complexes. Only EuO and YbO have become well-established because the $Eu^{2+}$ and $Yb^{2+}$ ions have a stable 4f electronic configuration without a 5d electron (i.e. half occupied $4f^7$ and fully occupied $4f^{14}$, respectively) [2].

Bulk polycrystals of *RE*Os (*RE* = La−Sm, Yb) have been synthesized from mixtures of *RE* metals and the oxides by high pressure synthesis in the early 1980s [8,9]. However, only electrical and magnetic properties of a small number of *RE*Os were reported [9,10], probably due to their chemical instability at ambient pressure [9] and the insufficient purity [11], as explained in more details [12]. The $RE^{2+}$ ions in *RE*Os mostly have a $4f^n5d^1$ electronic configuration (column 3 of Table 1), thus some of the light *RE*Os have been known to have electrical conduction in contrast with highly insulating $RE_2O_3$ [8]. Considering that stable rare earth nitrides *RE*N with $RE^{3+}$ ions are low temperature (anti)ferromagnetic insulators [13], *RE*Os except for EuO and YbO are expected to be high Curie temperature ($T_C$) **(see Glossary)** ferromagnetic conductors, since their itinerant 5d electrons could promote stronger carrier-mediated exchange coupling between localized spins of 4f electrons. However, many *RE*Os have not been synthesized yet, and the electronic and magnetic properties of most *RE*Os have not been investigated owing to the difficulty in their synthesis.

In 2016, rocksalt-type single crystalline YO was synthesized by thin film epitaxy **(see Glossary)** using pulsed laser deposition [14]. Thin film epitaxy enables the synthesis of various *RE*Os, some of which are novel solid-phase compounds. As expected, most *RE*Os show high electrical conduction, accompanied by a variety of magnetism properties, from superconductivity to room temperature ferromagnetism as a function of the *RE* element (i.e. f-electron number). In other words, nonmagnetic and insulating $RE_2O_3$ series are transformed into magnetic and highly conducting *RE*Os by subtracting one oxygen from the chemical formula $RE_2O_3$. In this review, the synthesis and fundamental electronic and magnetic properties of *RE*Os (*RE* = Y, La−Lu without Tm and radioactive Pm) are described [14−32]. So far, highly pure *RE*Os have not been systematically synthesized. Thus, this review is the first comprehensive review on the fundamental properties of *RE*Os. The *RE*Os in this review are either reborn or new compounds, whose synthesis needs non-equilibrium thin film growth under reductive conditions. The variety of fundamental properties despite the simple chemical formulae and crystal structure substantiates that "Less is also different"; a complementary term to P. W. Anderson's monumental words "More is different" [33,34]. In addition, the simple crystal structure is beneficial to tailor their heterostructures and superlattices like compound semiconductor devices. Hence, we hope this review would stimulate both experimental and theoretical studies on *RE*Os in the fields of not only pure science such as f-electron's solid state physics but also materials science, chemistry, and applications of the electrically conducting and magnetic f-electron's binary oxides for next generation electronic devices.

**Synthesis**

Pulsed laser deposition is a physical vapor deposition method to grow thin film on a substrate by evaporating solid-phase deposition source with a focused high power pulsed laser [35−37]. UV laser is frequently used for thin film growth of inorganic compounds, because high photon energy is necessary to evaporate inorganic compounds with the high melting temperatures. The frequently used lasers are excimer laser and Nd:YAG laser (with fourth harmonic wave), where the photon wavelength is 248 nm and 266 nm, respectively. When a single crystal substrate has a small lattice mismatch with the target compound, the epitaxial thin film can be obtained by optimizing the growth condition. Thin film growth process is a kinetic process to crystallize thin film instantly, in contrast with bulk crystal growth under thermal equilibrium conditions. Thus, it is possible to synthesize thermal non-equilibrium compounds, i.e. metastable compounds. Also, it is possible to grow a target compound selectively, e.g. among different polymorphs, by choosing a suitable single crystal substrate from the aspect of lattice matching between the target compound and the substrate.

For thin film growth of *RE*Os with anomalously low valent $RE^{2+}$ ions, highly reductive conditions are necessary. Hence, ultrahigh vacuum growth chamber is required with a limited supply of oxygen gas during thin film growth. As the deposition source, either *RE* metal or $RE_2O_3$ pellet is usually used because *RE*O bulk crystals are not easily available. However, the use of *RE* metal or $RE_2O_3$ often results in the formation of *RE* metal and/or $RE_2O_3$ impurities in thin film. Accordingly, it is important to optimize carefully the growth condition of thin film such as growth temperature (typically 100−500 °C) and oxygen pressure during growth (typically $1 \times 10^{-8}$−$1 \times 10^{-7}$ Torr). The choice of single crystal substrate commercially available is important, not only because the smaller lattice mismatch between the substrate and *RE*O stabilizes the *RE*O phase with less impurities, but also because the oxygen diffusion from oxide substrate during thin film growth could oxidize the *RE*O into $RE_2O_3$. In order to avoid such unfavorable oxidation, non-oxide substrates such as $CaF_2$ and oxide substrates with small oxygen diffusion such as $YAlO_3$ are often used. According to the lanthanoid contraction **(see Glossary)**, the lattice constant of *RE*Os decreases with increasing *RE*'s atomic number (column 4 of Table 1). Thus, the lattice mismatch between some *RE*O films and commercially available substrates is inevitably large. In order to reduce the lattice mismatch, the insertion of epitaxial buffer layer between *RE*O films and substrates is quite effective, resulting in the suppression of impurity phases. In addition, a capping layer on *RE*O film surface is necessary to avoid the film oxidation and degradation. Several nanometer-thick amorphous alumina capping layer in-situ grown at room temperature is very useful to prolong significantly the lifetime of *RE*O films, and such ultrathin capping layer enables surface sensitive photoemission spectroscopy of *RE*O films to characterize the electronic states. As a result, it is possible to obtain single crystalline and often impurity-free *RE*O films and to unveil various electronic and magnetic properties, as summarized (Table 1).

Table 1 Properties of rare earth monoxides.

| Compound | Impurity | Electronic config. of $RE^{2+}$ | Lattice const. [Å] | Electronic states | Carrier polarity | $\rho$ (300 K) [$\Omega$cm] | $n$ (300 K) [cm$^{-3}$] | $\mu$ (300 K) [cm$^2$/Vs] | Magnetism | Transition temp. [K] | $M_s$ (2 K) [$\mu_B$/f.u.] | Ref. |
|---|---|---|---|---|---|---|---|---|---|---|---|---|
| YO | Y$_2$O$_3$ | 4d$^1$ | $a$ = 4.936<br>$c$ = 4.977 | Semic# | n | 5.3×10$^{-3}$ | 1.9×10$^{21}$ | 0.42 | n/a | n/a | n/a | 14 |
| LaO | None | 5d$^1$ | $a$ = 5.198<br>$c$ = 5.295 | Metal | n | 1.0×10$^{-4}$ | 1.6×10$^{22}$ | 3.9 | SC | $T_c$ = 4.6 | n/a | 15 |
| CeO | None | 4f$^1$5d$^1$ | $a$ = 5.200<br>$c$ = 5.150 | Metal | n ($\leq$ 50 K)<br>p ($\leq$ 300 K) | 1.6×10$^{-4}$ | n: 1.4×10$^{22}$ (50 K)<br>p: 6.5×10$^{22}$ | n: 3.1 (50 K)<br>p: 0.54 | PM | n/a | n/a | 16 |
| PrO | None | 4f$^2$5d$^1$ | $a$ = 5.164<br>$c$ = 5.054 | Metal | n | 1.5×10$^{-4}$ | 2.9×10$^{22}$ (50 mK) | 2.3 (50 mK) | Weak FM | $T_C$ = 28 | 1.4 | 17 |
| NdO | None | 4f$^3$5d$^1$ | $a$ = 5.08/5.14<br>$c$ = 5.05/5.16 | Metal | n | 7.4×10$^{-5}$ | 1.1×10$^{23}$ | 0.77 | FM | $T_C$ = 19.1 | 1.1 | 18 |
| SmO | None | 4f$^5$5d$^1$ | $a$ = 4.96<br>$c$ = 5.02 | Metal | n | 1.6×10$^{-4}$ | 7.4×10$^{21}$ | 5.3 | PM | n/a | n/a | 19 |
| EuO | None | 4f$^7$ | $a$ = 5.143<br>$a$ = 5.134<br><br>$a$ = 5.116 | Semic | n | ~1×10$^8$<br>3.8×10$^{-1}$ | 2.5×10$^{19}$<br>~1×10$^{19}$ | 6.2 | FM | $T_C$ = 69<br>120<br>118<br>200 | 7<br>6.9<br>6.7 | 21<br>22<br>23<br>24 |
| GdO | Gd$_2$O$_3$ | 4f$^7$5d$^1$ | $a$ = 4.98<br>$c$ = 5.02 | Semic# | n | 8.4×10$^{-3}$<br>2.5×10$^{-3}$ | 1.7×10$^{21}$<br>1.1×10$^{21}$ | 0.44<br>2.4 | FM | $T_C$ = 276<br>303 | 1.3<br>0.87 | 25<br>26 |
| TbO | None | 4f$^8$5d$^1$ | $a$ = 4.90<br>$c$ = 5.03 | Metal | - | 9.5×10$^{-2}$ | - | - | FM | $T_C$ = 233 | 5.1 | 28 |
| DyO | None | 4f$^9$5d$^1$ | $a$ = 4.88<br>$c$ = 5.00 | Metal | - | - | - | - | FM | $T_C$ = 142 | 3.2 | 28 |
| HoO | Ho$_2$O$_3$ | 4f$^{10}$5d$^1$ | $a$ = 4.904<br>$c$ = 5.04 | Semic# | n | 3.50×10$^{-3}$ | 8.9×10$^{20}$ | 2 | FM | $T_C$ = 131 | 1.7 @5K | 29 |
| ErO | None | 4f$^{11}$5d$^1$ | $a$ = 4.84<br>$c$ = 4.97 | Metal | - | - | - | - | FM | $T_C$ = 88 | 5.6 | 28 |
| TmO | - | 4f$^{12}$5d$^1$ | - | - | - | - | - | - | - | - | - | |
| YbO | None | 4f$^{14}$ | $a$ = 4.83<br>$c$ = 4.87 | Semic | n | 2.2×10$^{-4}$ | 2.2×10$^{21}$ | 13 | n/a | n/a | n/a | 30 |
| LuO | Lu$_2$O$_3$ | 4f$^{14}$5d$^1$ | $a$ = 4.76<br>$c$ = 4.79 | Semic# | n | 1.8×10$^{-2}$ | 7.4×10$^{20}$ | 0.46 | n/a | n/a | n/a | 31 |

$\rho$: resistivity; $n$: carrier density; $\mu$: mobility; $M_s$: saturation magnetization; $T_c$: superconducting transition temperature; $T_C$: Curie temperature; Semic: semiconductor; SC: superconductor; PM: paramagnet; Ferro: ferromagnet; #: It could be metallic if rare earth sesquioxide impurity is eliminated.

**Structural properties**

The measured lattice constants of *RE*O films [14−19,25−31] and that of EuO bulk crystal [21] are summarized in Fig. 1a, as the most complete set of the *RE*O's lattice constants demonstrated for the first time. Crystal structure of the *RE*O films is tetragonally distorted rocksalt structure by epitaxial strain, where the *a*- and *c*-axis lengths are not equal. Thus, the cubic root of cell volume is also shown. From LaO to LuO, the lattice constant decreases almost monotonically, though the lattice constants of *RE*Os are generally larger than those of bulk *RE*Os [8], possibly caused by the presence of oxygen vacancy and/or the mixed valence of *RE* ion in *RE*O film [12]. Figure 1b shows ionic radii of the $RE^{2+}$ ions calculated from Fig. 1a as well as those of empirically calculated Shannon's $RE^{2+}$ ionic radii with six coordination [38,39]. These ionic radii show good coincidence each other. In the original Shannon's ionic radii table [38], only six-coordinated $Eu^{2+}$, $Dy^{2+}$, $Tm^{2+}$, and $Yb^{2+}$ ions are listed. Accordingly, Figure 1b provides a useful experimental dataset of six-coordinated $RE^{2+}$ ionic radii, contributing to structural characterizations and material designs of *RE* compounds with $RE^{2+}$.

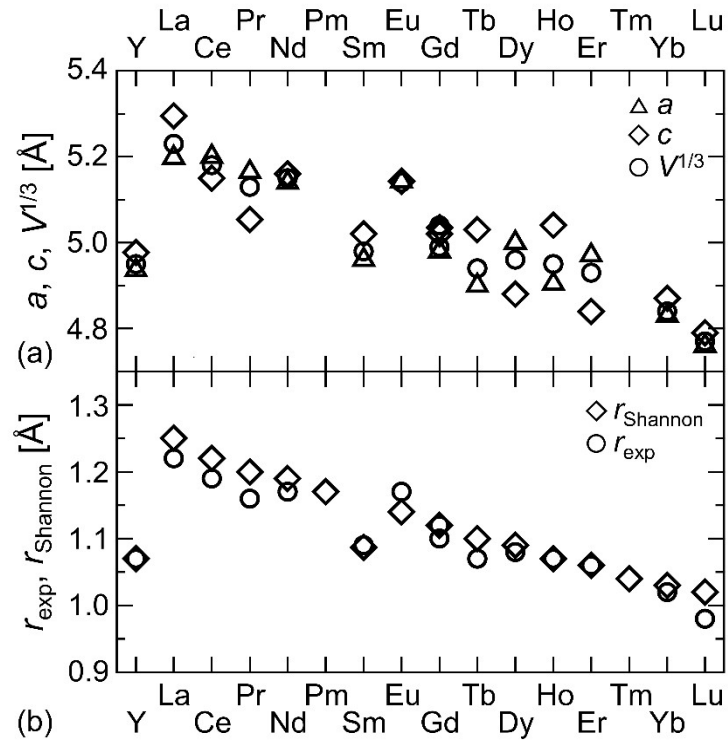

Figure 1. (a) In-plane and out-of-plane lattice constants, *a* and *c*, and cubic root of unit cell volume $V^{1/3}$ ($V = a^2c$) of rare earth monoxides (*RE*O) films [14−19,25,26,28−31]. (b) Ionic radii of the 6-coordinated $RE^{2+}$ ions obtained from the data ($V^{1/3}$) in (a), $r_{exp}$, and the Shannon's ionic radii assuming divalent rare earth ions [39].

**Electronic properties**

From electronic spectroscopies of *RE*Os [14,16,17,20,28,32], most of the $RE^{2+}$ ions possess $4f^n5d^1$ (n: 0−14) electronic configuration, except for $Y^{2+}$ ($4d^1$), $Eu^{2+}$ ($4f^7$), and $Yb^{2+}$ ($4f^{14}$) (column 3 of Table 1). Since 5d electrons are usually itinerant carrier, *RE* ions except for Eu and Yb ions in bulk *RE*Os are regarded to be in trivalent state as $RE^{3+}(O^{2-})(e^-)$ [8,11], while Sm ion is in a mixed valence state [10,20]. It is interesting whether the *RE* ions in *RE*O thin films are in divalent or mixed valence state, though the formal divalent state is used for the *RE* ions in this review.

All these *RE*Os show much higher electrical conduction in comparison with $RE_2O_3$ series as described below. Figure 2 shows the temperature dependence of resistivity **(see Glossary)** for *RE*O films. YO shows degenerate semiconductor behavior (Fig. 2a). The resistivity was influenced by oxygen pressure during growth [14], suggesting that the oxygen vacancy would serve as an electron donor like conventional oxide semiconductor. However, the YO film contained significant amount of $Y_2O_3$ impurity phase [14], which could suppress the electrical conduction in the YO thin film despite $4d^1$ electron configuration of $Y^{2+}$ ions. It is necessary to confirm the electrical conduction of YO by excluding $Y_2O_3$ impurity in the future.

LaO is a superconductor with the superconducting transition temperature ($T_c$) of about 5 K (Fig. 2b) [15]. La monochalcogenides except for LaO (LaS, LaSe, LaTe) have the higher $T_c$ for heavier chalcogen [40]. Contrary to this chemical trend, LaO has the highest $T_c$ among them, and the $T_c$ is varied by the amount of oxygen vacancies and the epitaxial strain [15].

CeO, PrO, NdO, and SmO show metallic conduction (Fig. 2b) [16−19]. CeO and SmO show the resistivity minimum probably originating from the Kondo effect **(see Glossary)**, often observed in heavy fermionic **(see Glossary)** Ce compounds [41]. Sm monochalcogenides except for SmO (SmS, SmSe, SmTe) are semiconducting at atmospheric pressure (called as black phase), and turn into metallic at high pressure (called as golden phase) [42,43]. In contrast, SmO is metallic even at atmospheric pressure. This result suggests that chemical pressure is generated in SmO due to the smallest O ion.

EuO with $4f^7$ electronic configuration is a well-known semiconductor [21]. Oxygen stoichiometric EuO is not highly electrically conducting, and turns to be highly electrically conducting by introducing oxygen vacancies or chemical dopants [22−24]. GdO with $4f^75d^1$ electronic configuration showed a semiconducting behavior in the first study [25]. However, GdO shows significantly lower resistivity with improved carrier mobility by reducing the amount of $Gd_2O_3$ impurity [26] (Fig. 2b). TbO [27], HoO [29], YbO [30], and LuO [31] also show semiconducting behavior (Figs. 2a and 2b), though TbO, HoO, and LuO contained non-negligible amounts of $RE_2O_3$ impurity phases. Recently, impurity-free TbO, DyO, and ErO films were obtained on lattice-constant-tunable (Ca,Sr)O buffer layers. These films show the clear Fermi edges, indicating metallic electronic states [28]. Taking into account these results, impurity-free YO, GdO, TbO, HoO, and LuO, that possess $4d^1$ or $4f^n5d^1$ electronic configuration, could show metallic conduction owing to itinerant nature of the d electron, though the semiconducting behaviors were reported for those containing $RE_2O_3$ impurities. YbO would be intrinsically semiconducting because of the $4f^{14}$ electronic configuration.

From the Hall effect **(see Glossary)** measurements, most *RE*Os show n-type conduction (column 6 of Table 1), though the Hall effect of TbO, DyO, ErO, and TmO has not been reported yet. Only CeO shows p-type conduction below 300 K, where the hole mobility is largely enhanced to 646 $cm^2V^{-1}s^{-1}$ below 5 K accompanied with n-type conduction [16]. This result is consistent with a band calculation study, where both 4f and 5d states reside near the Fermi energy with a hole pocket [44]. The electronic properties of all reported *RE*Os are summarized (columns 5−9 of Table 1).

While the room temperature resistivity of chemically stable $RE_2O_3$ is the order of $10^{10}$ Ω·cm or higher [2], that of *RE*Os is extraordinarily smaller, the order of $10^{-4}$–$10^{-2}$ Ω·cm, owing to the additional $4d^1$ or $5d^1$ electrons in *RE*Os. Indeed, such high electrical conduction has not been reported in $RE^{2+}$ metal complexes [7], due to the spatially isolated $RE^{2+}$ ions and $5d^1$ electrons in the large molecular structure in contrast with the densely packed $RE^{2+}$ ions in rocksalt-type *RE*Os. Although the electrical conduction of *RE*Os is similar (Fig. 2), their details are different: superconducting or not, p-type or n-type conduction, etc. This fact means that the 4f electrons, which are localized near inner shell, significantly influence the electronic states of *RE*Os.

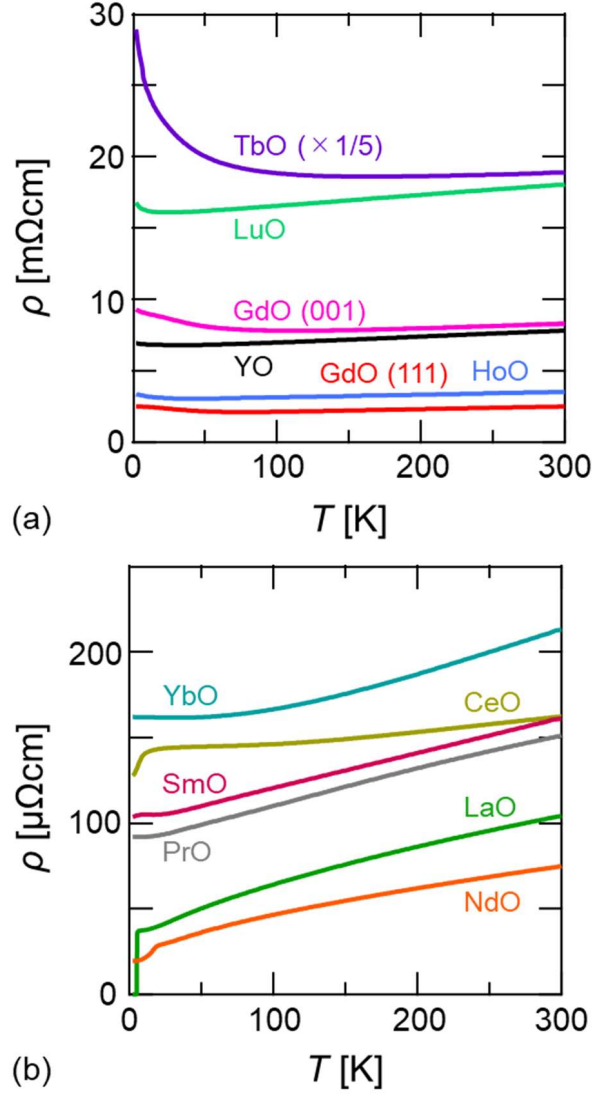

Figure 2. Temperature dependence of resistivity for rare earth monoxide (*RE*O) films (a) with rare earth sesquioxide (*RE*$_2$O$_3$) impurity and (b) without *RE*$_2$O$_3$ impurity [14−19,25−27,29−31].

**Magnetic properties**

In contrast with *RE*$_2$O$_3$ series, magnetism in *RE*Os is significantly influenced by the f-electron's number, as shown in Fig. 3.

For light *RE*Os (*RE* = La−Sm), LaO is superconducting below 5 K (Fig. 3a) [15], CeO is paramagnetic (Fig. 3b) [16], while PrO is weakly ferromagnetic with the $T_C$ of 28 K (Fig. 3c) [17]. NdO is ferromagnetic with the $T_C$ of 19 K (Fig. 3d) [18], but its ultrathin film shows small but finite magnetic hysteresis up to the doubled $T_C$, 40 K, as described later [32]. SmO is paramagnetic [19]. EuO has been well-known n-type ferromagnetic semiconductor with the $T_C$ of 69 K [22], where the $T_C$ is largely enhanced up to 200 K by electron doping [23,24].

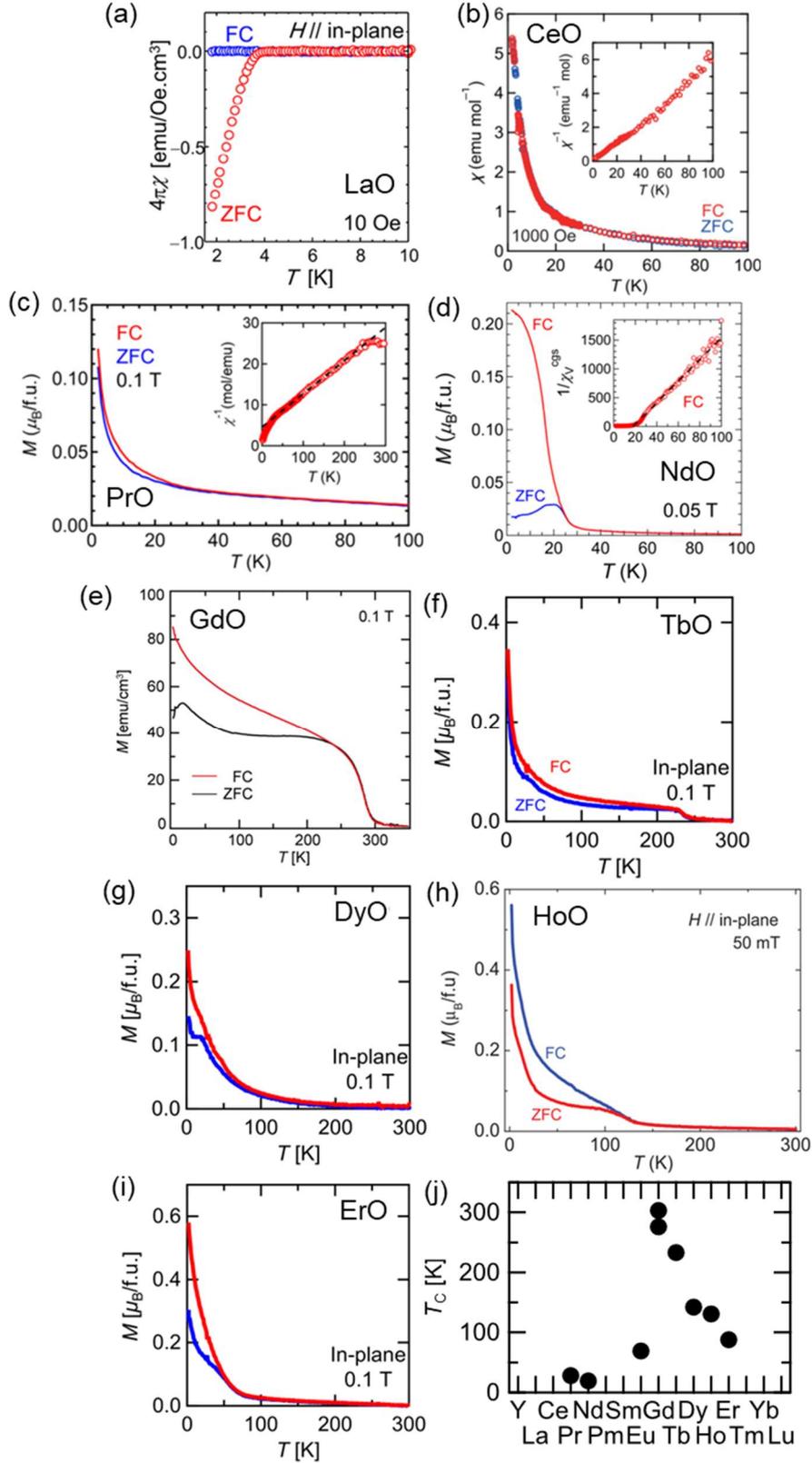

Figure 3. (a)–(i) Temperature dependence of magnetization for rare earth monoxide (REO) films. (j) The Curie temperature ($T_C$) of REO films. (a) LaO. (Ref. 15. ©2018 American Chemical Society.) (b) CeO. (Ref. 16. ©2022 American Physical Society.) (c) PrO. (Ref. 17. ©2022 American Physical Society.) (d) NdO. (Ref. 18. ©2019 American Physical Society.) (e) GdO. (Ref. 26. ©Royal Chemical

Society.) (f) TbO. (Ref. 28. ©Royal Chemical Society.) (g) DyO (Ref. 28. ©Royal Chemical Society.) (h) HoO. (Ref. 29. ©AIP Publishing.) (i) ErO. (Ref. 28. ©Royal Chemical Society.) (j) $T_C$ of *RE*Os as a function of rare earth element.

For heavy *RE*Os (*RE* = Gd−Lu), GdO is ferromagnetic with the $T_C$ of 276 K, that is significantly higher than the $T_C$ of EuO [25]. The $T_C$ of GdO is still enhanced up to 303 K (Fig. 3e), by the improved crystallinity and/or the introduction of oxygen vacancy [26]. TbO, DyO, HoO, and ErO are ferromagnetic with the $T_C$ of 233 K (Fig. 3f) [27,28], 142 K (Fig. 3g) [28], 131 K (Fig. 3h) [29], and 88 K (Fig. 3i) [28], respectively. YbO with fully occupied $4f^{14}$ [30] and LuO with $4f^{14}5d^1$ [31] are not ferromagnetic but probably diamagnetic and paramagnetic, respectively. Magnetic properties of all reported *RE*Os are summarized (columns 10−12 of Table 1).

Rare earth monopnictides *REPn*s (*Pn* = N, P, As, Sb, Bi) with the $RE^{3+}$ ions and rare earth monochalcogenides *RECh*s (*Ch* = S, Se, Te) with the $RE^{2+}$ ions have been reported, and their magnetism has been already known [45,46]. For *REPn*s and *RECh*s (*RE* = Nd, Eu, Gd) except for paramagnetic Eu*Pn*s, the $T_N$ is higher for heavier *Pn* and *Ch*, while the higher $T_C$ than the $T_N$ appears for the lightest N and O [18,25]. In other words, the ferromagnetism dominates over the antiferromagnetism accompanied with the higher $T_C$ for the shorter nearest-neighbor *RE*−*RE* distance. Taking into account that each *RE*O tends to have higher $T_C$ than each *RE*N, the 5d electrons in *RE*Os probably promote carrier-mediated exchange interaction between the *RE* ions. It is well known that EuO shows significantly enhanced $T_C$ by electron doping. On the other hand, it is uncertain if such electron doping increases the $T_C$ of *RE*Os, because of the intrinsically high carrier density of the *RE*Os (*RE* ≠ Eu, Yb) unlike EuO [21]. Indeed, La-doped EuO showed highly improved electrical conduction with triply enhanced $T_C$ [24], while La-doped GdO showed no enhancement of the electron density and $T_C$ [25]. Figure 3j shows the $T_C$ of *RE*Os. Only PrO and NdO are ferromagnetic among the light *RE*Os with the $T_C$ below 30 K. On the other hand, many *RE*Os are ferromagnetic among the heavy *RE*Os with the range of $T_C$ from 100 to 300 K, exhibiting monotonic decreasing $T_C$ with increasing the f-electron's number.

As described above, *RE*Os have the highest $T_C$ among the corresponding *REPn*s and *RECh*s, while $RE^{2+}$ metal complexes show no ferromagnetic order probably due to the spatially isolated $RE^{2+}$ ions [47]. On the other hand, the saturation magnetization of *RE*Os is significantly smaller than that of free *RE* ions, similar to *RE*N [13,48], possibly caused by effects of crystal field under lattice strain in the film [13], non-collinear spin structure [49], and/or antiferromagnetic superexchange coupling via the oxygen ions. Such non-collinear spin structure observed in Ho*Pn*s [50−52] would also appear in HoO, which exhibited a step-like magnetic hysteresis just below the $T_C$ [29], though such spin structure in HoO would be preserved at much higher temperature than other Ho*Pn*s because of the much higher $T_C$ of HoO.

The temperature dependence of magnetization for GdO, TbO, DyO, HoO, and ErO is not conventional convex shape but concave shape at low temperatures (Figs. 3e−i). Such temperature dependence was proposed to appear in magnetic semiconductors owing to the formation of bound magnetic polaron [53,54]. Another possible explanation would be spin reorientation of 4f spin below the $T_C$, as was observed in rare earth metals and compounds [55], though the possibility of insufficient crystallinity of the *RE*O films cannot be ruled out.

The other remarkable feature of *RE*Os is magnetic anisotropy (**see Glossary**). The *RE*O films have cubic or tetragonally distorted rocksalt structure, thus the magnetocrystalline anisotropy is almost cubic. However, shape magnetic anisotropy often governs the magnetic anisotropy in magnetic thin films in case of the large magnetization. Hence, EuO film shows a large shape magnetic anisotropy, where the easy axis is along the in-plane [22]. However, *RE*O films except for EuO film usually show insignificant shape magnetic anisotropy but approximately cubic magnetic anisotropy, exhibiting insignificant difference between the in-plane and out-of-plane magnetization curves [27]. Such cubic magnetic anisotropy is beneficial to realize perpendicular magnetization under small external magnetic field, enabling to observe the anomalous Hall effect clearly as described below.

**Anomalous Hall effect**

Magnetic *RE*Os are highly conducting, thus show the anomalous Hall effect (**see Glossary**). Figure 4 shows the magnetic field dependence of anomalous Hall effect for *RE*Os. CeO shows the magnetic-field-nonlinear paramagnetic anomalous Hall effect (Fig. 4a) [16]. PrO and NdO show the ferromagnetic anomalous Hall effect, that is positively proportional to the magnetization, i.e. the positive anomalous Hall coefficients (Figs. 4b and 4c) [17,18]. Meanwhile, GdO shows the ferromagnetic anomalous Hall effect with negative anomalous Hall coefficient (Fig. 4d) [26], similar to EuO [22]. Subsequently, HoO shows the ferromagnetic anomalous Hall effect with positive anomalous Hall coefficient as well as step-like magnetic field dependence (Figure 4e), as was observed in the magnetization curve [29]. Some of these *RE*Os show apparently different magnetic field dependences between the anomalous Hall effect and magnetization: e.g. the significantly larger remanence (**see Glossary**) and coercive field (**see Glossary**) for the former in PrO (Fig. 4b) [17]. Such difference suggests non-collinear spin structure of these *RE*Os as was reported in other antiferromagnetic compounds with non-collinear spin structure [56−60].

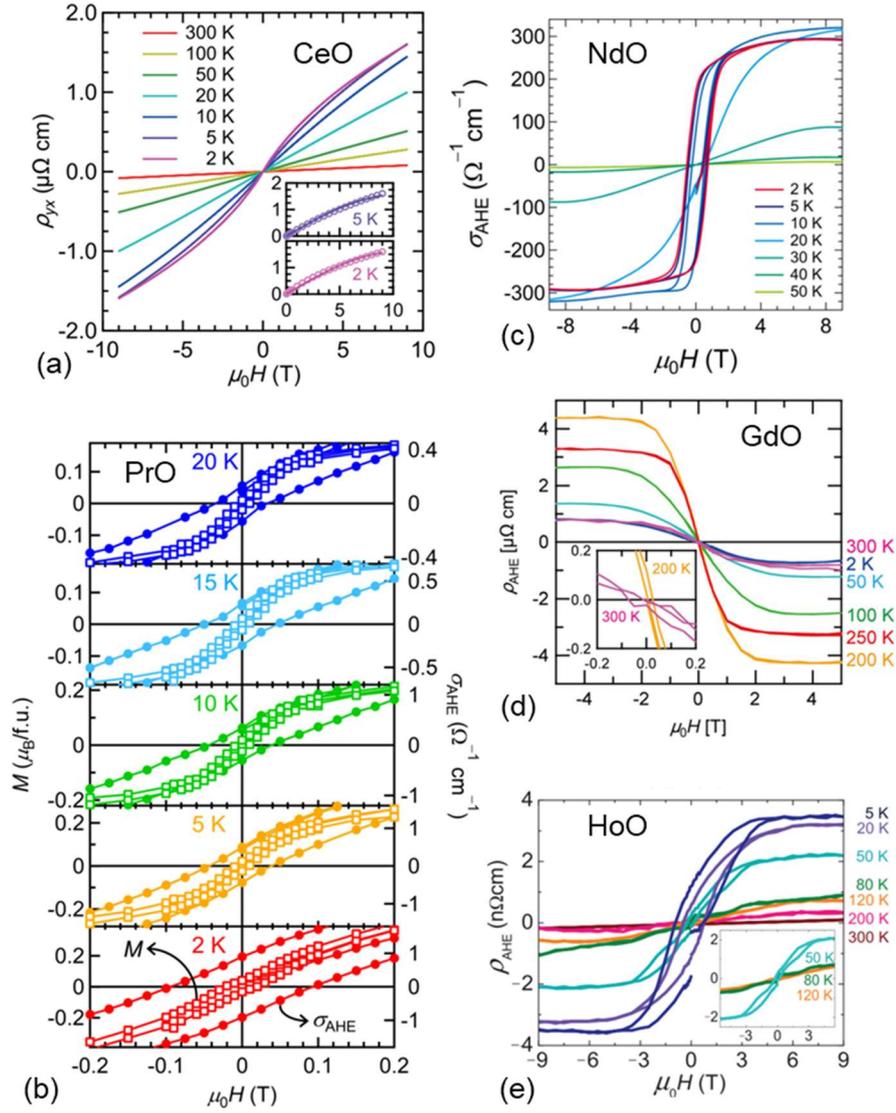

Figure 4. Anomalous Hall effect of rare earth monoxide (*RE*O) films. (a) CeO (Ref. 16. ©2022 American Physical Society.) (b) PrO (Ref. 17. ©2022 American Physical Society.) (c) NdO (Ref. 18. ©2019 American Physical Society.) (d) GdO (Ref. 26. ©Royal Chemical Society.) (e) HoO (Ref. 29. ©AIP Publishing.)

The anomalous Hall effect measurement is not difficult even for ultrathin films, in contrast with the difficulty in magnetization measurement due to the small magnetization signal. Thus, the anomalous Hall effect can be a useful electrical probe for magnetization. Generally, the $T_C$, magnetization, and electrical conductivity are monotonic decreasing with decreasing film thickness for ferromagnetic ultrathin films. However, NdO films show different behavior. The 12 nm-thick NdO film shows small but finite hysteretic anomalous Hall effect up to the doubled $T_C$, 40 K, indicating the presence of weakly ferromagnetic order at such high temperature (Fig. 5) [32]. Such ferromagnetic order above $T_C$ was observed in Gd and Tb metals [61,62], attributed to the interaction between 4f and itinerant electrons [63,64].

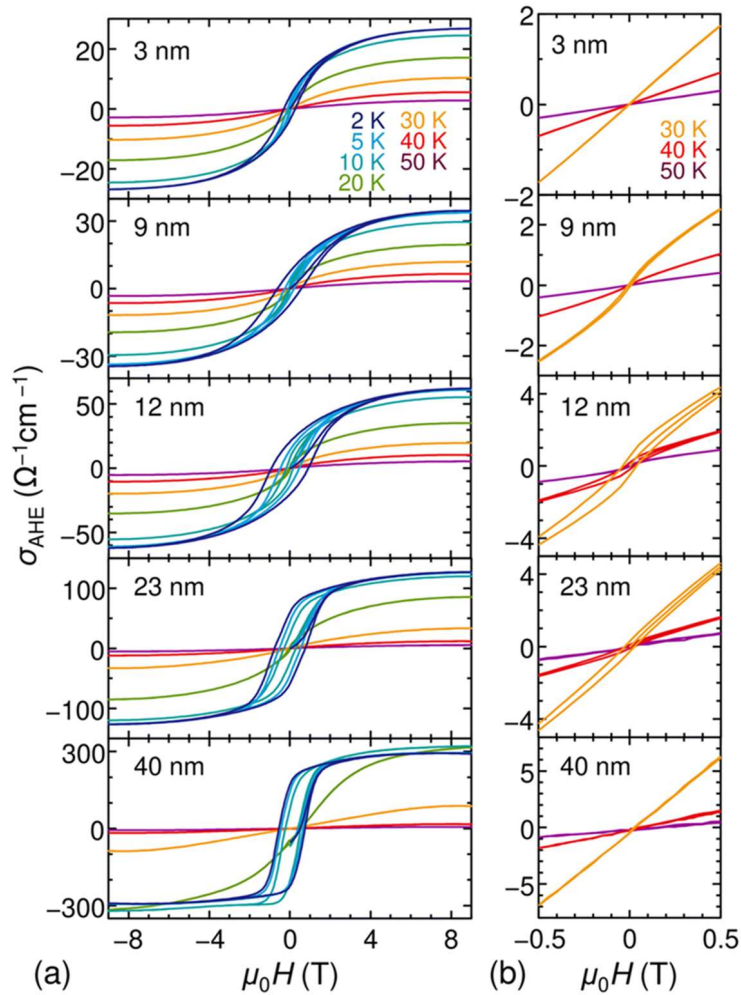

Figure 5. (a) Thickness dependence of anomalous Hall conductivity for NdO films. (b) The magnified data at 30, 40, and 50 K. (Ref. 32. ©Royal Chemical Society.)

**Rocksalt heteroepitaxial structure**

*RE*Os have a very simple rocksalt structure, thus it would be easy to combine different *RE*O layers coherently as the heteroepitaxial structures. Indeed, superconducting LaO layer was epitaxially grown on ferromagnetic EuO layer [65]. In the rocksalt-type LaO/EuO heteroepitaxial bilayer, the superconductivity of LaO was suppressed by only 2−6 nm-thick EuO. In previous studies on superconductor/ferromagnet heterostructures, the superconductivity was not fully suppressed in Al/EuS (5 nm) bilayer [66], while was suppressed in $YBa_2Cu_3O_{7−\delta}$/$La_{2/3}Ca_{1/3}MnO_3$ (13 nm) hetroepitaxial bilayer [67]. These results suggest that the sufficiently sharp LaO/EuO interface was formed to fully suppress the superconductivity, because of the same crystal structure of LaO and EuO.

At present, no other study on *RE*O heterostructures has not been reported except for theoretical calculation of SmO/EuO interface, that was proposed to exhibit quantum anomalous Hall effect [68]. On the other hand, several *RE*N heterostructures were reported to exhibit the spin-spin interaction between the adjacent layers [69,70]. The higher electrical conduction of *RE*O than that of *RE*N would enable the electrical detection of such spin-spin interaction. Because of many possible combinations of *RE*Os in the heterostructures, theoretical prediction of physical properties would be very helpful for the future studies.

**Concluding Remarks**

Rare earth binary oxides are usually insulating and either nonmagnetic or low temperature antiferromagnetic. However, high electrical conduction, ferromagnetism, superconductivity, or heavy fermion behavior are unveiled in rare earth monoxides. The rare earth monoxides have a simple rocksalt structure and chemical composition, though most of them have been chemically unstable owing to their unusual divalent rare earth ions. By simply using ultraviolet laser ablation of rare metals or oxides in very diluted oxygen atmosphere, almost full set of single crystalline rare earth monoxides is synthesized in a form of epitaxial thin film, and a variety of electronic and magnetic properties are unveiled. Their highly pure bulk synthesis would expand research fields of rare earth monoxides. It is intriguing that magnetism in the rare earth monoxides strongly depends on the f-electron's number despite the spatially localized nature of the f-electron near inner shell, in contrast with chemically stable rare earth sesquioxides. It is also noted that well-known rocksalt-type 3d transition-metal monoxides are semiconducting and antiferromagnetic [71]. Accordingly, high electrical conduction and high temperature ferromagnetism in the rare earth monoxides are their unique properties among all rocksalt-type metal oxides. In general, the magnetization in transition metal compounds originates from spin angular moments, while that in rare earth monoxides originates from spin and orbital angular moments. Thus, heavy rare earth monoxides have large magnetization because of the parallel coupling between spin and orbital angular moments. In addition, the large spin-orbit interaction is expected because of the heavy mass of rare earths. So far, f-electron systems have been studied from the aspect of pure science, but the high electrical conduction and ferromagnetism are worth investigating from the aspect of applications such as electronics, magnetics, and spintronics. Additionally, it is interesting whether the unusual divalent state is useful or not for chemical applications such as electrocatalysts.

Recently, rare earth monoxides including the bulk crystals revive in experimental studies [12,72−73]. However, the theoretical and calculational studies are limited to LaO [74−76]. As described in this review, fourteen rare earth monoxides are available at least. The rocksalt-type rare earth monoxides can be regarded as Lego blocks for heterostructures and superlattices to implement electronic, spintronic, superconducting, and topological functionalities promising for next generation electronic devices. The simple crystal structure and chemical composition would contribute to not only producing novel or exotic properties in heterostructures and heterointerfaces but also further understanding and new paradigm of f-electron systems. Finally, it would be possible to discover novel electronic and magnetic properties in unexplored metastable solids with the unusual valence ions of transition metal and actinoid elements by developing suitable synthetic methods. Such discovery will pave the way for new fields of science and technology.

**Acknowledgements**

This work is supported by JSPS-KAKENHI (Grant No. 26105002, 18H03872, 18K18935, 21H05008), JST-CREST (Grant No. JPMJCR1173), and the Mitsubishi Foundation. We acknowledge Hiroshi Kumigashira and Daisuke Shiga for useful discussions.

**Glossary**

**Coercive field**: External magnetic field required to force zero magnetization in ferromagnets, i.e. $x$-intercept in magnetic hysteresis curve.

**Curie temperature**: Magnetic transition temperature of ferromagnets, below which ferromagnetism appears.

**Field effect transistor**: A transistor exhibiting tunable electrical conduction with the application of electric field.

**Hall effect**: Ordinary Hall effect is an electromotive force perpendicular to electric current and magnetic field in nonmagnetic conductor. Anomalous Hall effect is an electromotive force perpendicular to electric current and magnetization in magnetic conductor. The anomalous Hall effect is usually superposed with the ordinary Hall effect in the Hall effect measurement.

**Heavy fermion**: Heavy effective mass of electrons (or holes) in conductors, which usually contain heavy element.

**Kondo effect**: The scattering of electron carriers by magnetic impurities in conductors, generating the resistivity minimum followed by resistivity upturn with decreasing temperature.

**Lanthanoid contraction**: The shrinkage of ionic and atomic radii with the atomic number for lanthanoid elements.

**Magnetic anisotropy**: The tendency of magnetization in ferromagnets to point to a specific direction such as crystallographic axis.

**Néel temperature**: Magnetic transition temperature of antiferromagnets, below which antiferromagnetism appears.

**Remanence**: Residual magnetization in ferromagnets under zero magnetic field, i.e. $y$-intercept in magnetic hysteresis curve.

**Resistivity**: The inverse of electrical conductivity that is a quantitative measure of electrical conduction.

**Thin film epitaxy**: The growth of single crystalline thin film on single crystal substrate, maintaining a definite crystallographic orientation between the thin film and the substrate.